\documentclass[aps,prd,twocolumn,superscriptaddress,tightenlines,nofootinbib]{revtex4-1}
\usepackage[T1]{fontenc}
\usepackage{microtype}
\usepackage[textsize=scriptsize,backgroundcolor=red!70,linecolor=red]{todonotes}
\usepackage{booktabs}
\usepackage{dcolumn}
\usepackage{amsmath}
\usepackage{amsfonts}
\usepackage{amssymb}
\usepackage{graphicx}
\usepackage{hyperref}
\usepackage[all]{hypcap}
\usepackage{paralist}
\usepackage{multirow}

\newcommand{\chisq}{\ensuremath{\chi^2}}

\usepackage{graphicx}
\usepackage{dcolumn}
\usepackage{bm}
\usepackage{epsf}
\usepackage{dcolumn}
\def\be{\begin{equation}}
\def\ee{\end{equation}}
\def\bea{\begin{eqnarray}}
\def\eea{\end{eqnarray}}
\def\gsim{\ \rlap{\raise 2pt\hbox{$>$}}{\lower 2pt \hbox{$\sim$}}\ }
\def\lsim{\ \rlap{\raise 2pt\hbox{$<$}}{\lower 2pt \hbox{$\sim$}}\ }
\def\dslash{\kern-4pt \not{\hbox{\kern-2pt $\partial$}}}
\def\pslash{\not{\hbox{\kern-2pt p}}}

\begin{document}
\DeclareGraphicsExtensions{.eps,.ps}


\title{MiniBooNE, MINOS+ and IceCube data imply a baroque neutrino sector}



\author{Jiajun Liao}
\affiliation{Department of Physics and Astronomy, University of Hawaii at Manoa, Honolulu, HI 96822, USA}
\affiliation{School of Physics, Sun Yat-Sen University, Guangzhou 510275, China}

\author{Danny Marfatia}
\affiliation{Department of Physics and Astronomy, University of Hawaii at Manoa, Honolulu, HI 96822, USA}

\author{Kerry Whisnant}
\affiliation{Department of Physics and Astronomy, Iowa State University, Ames, IA 50011, USA}

\begin{abstract}
The 4.8$\sigma$ anomaly in MiniBooNE data cannot be reconciled with MINOS+ and IceCube data within the vanilla framework of neutrino oscillations involving an eV-mass sterile neutrino. We show that an apparently consistent picture can be drawn if charged-current and neutral-current nonstandard neutrino interactions are at work in the 3+1 neutrino scheme. It appears that either the neutrino sector is more elaborate than usually envisioned, or one or more datasets needs revision. 
  
\end{abstract}
\maketitle

{\bf Introduction.}
The existence of an eV scale neutrino has been a major open question in neutrino physics for more than two decades. 
Recently, the MiniBooNE collaboration updated their analysis after 15 years of running, and reported a 4.8$\sigma$~C.L. excess in the electron and anti-electron neutrino spectra close to the experimental threshold~\cite{Aguilar-Arevalo:2018gpe}. An explanation of the results via 
       $\nu_\mu\rightarrow\nu_e$ oscillation with a mass-squared difference $\delta m^2\sim 1 \text{ eV}^2$, is consistent with the LSND anomaly found at a similar $L/E\sim 1$ m/MeV~\cite{Aguilar:2001ty}. The two excesses combined have reached a significance of 6.1$\sigma$~C.L., and urgently call for an explanation that makes them compatible with other experiments.
       
\begin{figure}
	\includegraphics[width=0.45\textwidth]{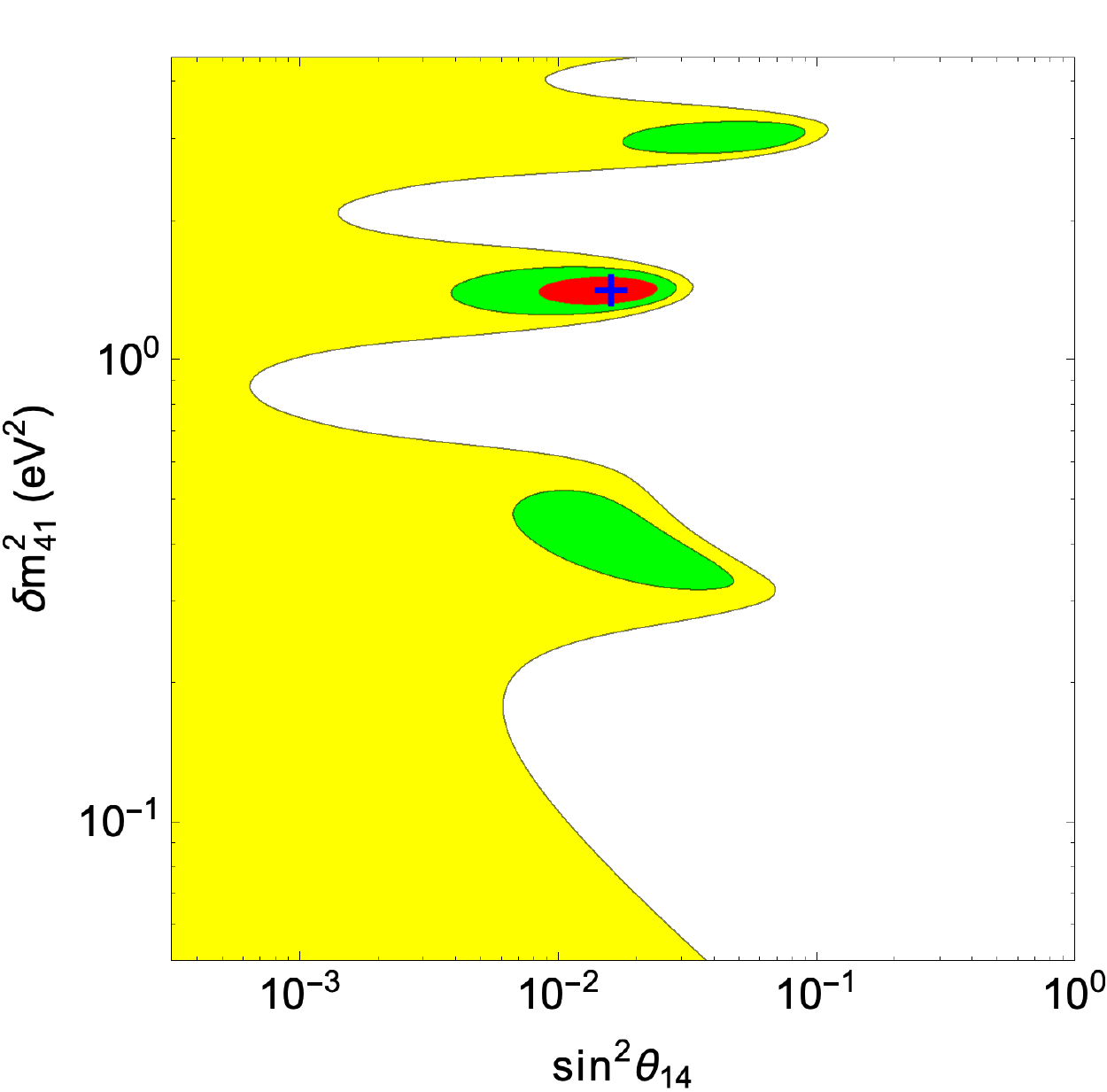}
	\caption{The $1\sigma$, $2\sigma$, and $3\sigma$ regions allowed by DANSS. The blue  plus sign marks the best fit point, $\delta m_{41}^2=1.4\text{ eV}^2$ and $\sin^2 \theta_{14}=0.016$.}
	\label{fig:danss}
\end{figure}

It is well known that the appearance data are in serious tension with disappearance data in global fits of the 3+1 oscillation framework~\cite{Collin:2016rao, Gariazzo:2017fdh, Dentler:2018sju}. To explain the LSND and MiniBooNE excess via sterile neutrino oscillations, a relatively large mixing amplitude $\sin^2 2\theta_{\mu e}\equiv4|U_{e4}U_{\mu4}|^2=\sin^2 2\theta_{14}\sin^2\theta_{24}$ is required. Constraints on $|U_{e4}|=\sin\theta_{14}$ are provided mainly by reactor neutrino experiments, with Daya Bay contributing a strong constraint on $|U_{e4}|$ for $\delta m_{41}^2<0.5 \text{ eV}^2$~\cite{An:2016luf}.  Interestingly, recent fits to data from the reactor experiments, NEOS~\cite{Ko:2016owz} and DANSS~\cite{Alekseev:2018efk}, suggest a sterile neutrino interpretation at the $3\sigma$ level with $\delta m_{41}^2\approx1.3 \text{ eV}^2$ and $|U_{e4}|^2\approx0.01$~\cite{Dentler:2018sju, Gariazzo:2018mwd}. In particular, since the DANSS experiment measured the ratios of energy spectra at different distances, the results are independent of the uncertain reactor $\bar{\nu}_e$ flux. Our analysis of the measured bottom/top ratios of the positron energy spectra in Ref.~\cite{Alekseev:2018efk} gives the best-fit parameters, $\delta m_{41}^2=1.4\text{ eV}^2$, $\sin^2 \theta_{14}=0.016$ with a $\chisq$ value smaller by 10.8 than for the standard no oscillation case. The regions favored by DANSS are shown in Fig.~\ref{fig:danss}.  We see that DANSS data prefer an eV-scale neutrino oscillation with the mixing angle $\sin^2 \theta_{14}$ in the $1\sigma$ range, $0.0087-0.023$. Our results are consistent with those of Refs.~\cite{Dentler:2018sju, Gariazzo:2018mwd} after taking into account the large systematic uncertainties due to the energy resolution and the sizes of the source and detector.

       Constraints on $|U_{\mu4}|=\cos\theta_{14}\sin\theta_{24}$, which are mostly driven by the $\nu_\mu$ disappearance experiments at IceCube~\cite{TheIceCube:2016oqi} and MINOS+~\cite{Adamson:2017uda}, rule out the 3+1 scenario for the MiniBooNE/LSND data. Hence, if we take the results of all three experiments, MiniBooNE, MINOS+ and IceCube, at face value, a {\it baroque} new physics scenario must be introduced to explain all the data. In this Letter, we first show that the MINOS+ constraints can be relaxed if there exist charged-current (CC) nonstandard interactions (NSI) in the detector. (An earlier analysis invoked CC NSI in the 3+1 scenario  to explain a discrepancy between neutrino and antineutrino oscillations observed in early MiniBooNE data~\cite{Akhmedov:2010vy}.) It is known that large neutral current (NC) NSI, can suppress the resonant enhancement of high energy atmospheric neutrino oscillations, and weaken the IceCube constraints on 
       $\sin\theta_{24}$~\cite{Liao:2016reh}. Since large NC NSI also modify the lower energy atmospheric neutrino spectrum at DeepCore, here we study NSI effects on the combination of IceCube and DeepCore data.

{\bf Framework.}
We consider the simplest 3+1 mass scheme, with an eV-mass sterile neutrino in addition to the three active neutrinos. CC and NC NSI are motived by new physics beyond the standard model, and their effects on neutrino oscillations have been extensively studied; for reviews see Ref.~\cite{Ohlsson:2012kf, Miranda:2015dra, Farzan:2017xzy}. Similar to the standard electroweak interactions, the NSI we require can be described by the dimension-six operators, 
\begin{align}
\mathcal{L}_\text{NC-NSI} &=-2\sqrt{2}G_F
\epsilon^{f C}_{\alpha\beta} \!
\left[ \overline{\nu_\alpha} \gamma^{\rho} P_L \nu_\beta \right] 
\left[ \bar{f} \gamma_{\rho} P_C f \right]\,,
\\
\mathcal{L}_\text{CC-NSI} &=-2\sqrt{2}G_F
\epsilon^{ff^\prime C}_{\alpha\beta} \!
\left[ \overline{\nu_\beta} \gamma^{\rho} P_L \ell_\alpha \right] 
\left[ \bar{f^\prime} \gamma_{\rho} P_C f \right]\,,
\label{eq:NSI}
\end{align}
where $\alpha, \beta\in {e, \mu, \tau, s}$, $C=L,R$, $f \neq f^\prime\in {u,d}$, $f,\, f^\prime\neq e$, and $\epsilon^{fC}_{\alpha\beta}$ and $\epsilon^{ff^\prime C}_{\alpha\beta}$ are dimensionless and parmeterize the
strength of the new interactions in units of the Fermi constant $G_F$.
The NC NSI mainly affect neutrino propagation in matter, and the CC NSI affect neutrino production and detection. 
Hence, when both NC and CC NSI are operative, the apparent oscillation probability measured in an experiment can be written as~\cite{Kopp:2006wp}
\begin{align}
\tilde{P}(\nu_\alpha^S\rightarrow\nu_\beta^D)=\left|\left[(1+\epsilon^D)^Te^{-iHL}(1+\epsilon^S)^T\right]_{\beta\alpha}\right|^2\,,
\label{eq:S}
\end{align}
where $\epsilon^S_{\alpha\beta}$ and $\epsilon^D_{\alpha\beta}$ are defined through the CC NSI parameters $\epsilon^{ff^\prime C}_{\alpha\beta}$, and the Hamiltonian $H$ is given by
\be
H = {1\over2E} \left[  V
\left( \begin{array}{cccc}
	0 & 0 & 0 & 0\\ 0 & \delta m^2_{21} & 0 & 0\\ 0 & 0 & \delta m^2_{31} & 0\\ 0 & 0 & 0 & \delta m^2_{41}
\end{array} \right)
V^\dagger \right]+V_\text{m}\,,
\label{eq:H}
\ee
with $\delta m_{ij}^2=m_i^2-m_j^2$, and $V=R_{34}O_{24}O_{14}R_{23}O_{13}R_{12}$. Here $R_{ij}$ is a real rotation by an angle $\theta_{ij}$ in the $ij$ plane, and $O_{ij}$ is a complex rotation by $\theta_{ij}$ and a phase $\delta_{ij}$. 
The matter potential in Eq.~(\ref{eq:H}) is 
\begin{align}
V_m=V_{CC} \begin{pmatrix}
1+\epsilon_{ee}^m & \epsilon_{e\mu}^m & \epsilon_{e\tau}^m & \epsilon_{e s}^m \\
\epsilon_{e\mu}^{m*} & \epsilon_{\mu\mu}^m & \epsilon_{\mu\tau}^m & \epsilon_{\mu s}^m \\
\epsilon_{e\tau}^{m*} &\epsilon_{\mu\tau}^{m*} & \epsilon_{\tau\tau}^m & \epsilon_{\tau s}^m \\
\epsilon_{e s}^{m*} &\epsilon_{\mu s}^{m*} & \epsilon_{\tau s}^{m*} & \kappa+\epsilon_{s s}^m  \\
\end{pmatrix}  \,,
\end{align}
where $V_{CC}=\sqrt{2}G_FN_e$ is the electron charged-current potential, $\kappa=\frac{N_n}{2N_e}\simeq 0.5$ is the standard NC/CC ratio, and $\epsilon_{\alpha\beta}^m\equiv\sum\limits_{f,C}\epsilon^{fC}_{\alpha\beta}\frac{N_f}{N_e}$ is the effective strength of NSI in matter, and $N_f$ is the number density of fermion $f$. 


        To relax the MINOS+ and IceCube bounds, we employ Occam's razor and assume that only $\epsilon_{\mu\mu}^{D}$, 
        $\epsilon_{\mu\mu}^m$,  $\epsilon_{\tau\tau}^m$ and $\epsilon_{ss}^m$ are nonzero. Note that CC NSI at the source may be different from those at the detector. Since neutrinos produced by the NuMI beamline mainly arise from pion decay and pions only couple to the axial-vector current, a vector-like interaction, i.e., $\epsilon^{udL}_{\mu\mu}=\epsilon^{udR}_{\mu\mu}$, only yields CC NSI at the detector~\cite{Akhmedov:2010vy}. Then, $\epsilon^{D}_{\mu\mu} = 2\epsilon^{udL}_{\mu\mu}$. We set $\theta_{34}$ and all phases to be equal to zero for simplicity. 
Since the $\nu_e$ flux is small compared to the $\nu_\mu$ flux at these experiments, and the $\nu_e$ mixing is suppressed by $s_{13}^2$ and $s_{14}^2$, we ignore the $\nu_e$ component and consider a three flavor system with only $\nu_\mu$, $\nu_\tau$, and $\nu_s$. After a rotation by $R_{24}$, the Hamiltonian that describes the three neutrino propagation in matter can be written as
\begin{align}
&R_{24}^\dagger H R_{24} \approx {\delta m_{31}^2\over2E}   \times
\nonumber\\
&{\tiny \left( \begin{array}{ccc}
	 s_{23}^2+\hat{A}(c_{24}^2\epsilon_{\mu\mu}^m+s_{24}^2\tilde{\epsilon}_{ss}^m) & c_{23}s_{23} & \hat{A}c_{24}s_{24}(\epsilon_{\mu\mu}^m-\tilde{\epsilon}_{ss}^m)\\ c_{23}s_{23} & c_{23}^2+\hat{A}\epsilon_{\tau\tau}^m & 0\\  \hat{A}c_{24}s_{24}(\epsilon_{\mu\mu}^m-\tilde{\epsilon}_{ss}^m) & 0 & R+\hat{A}(c_{24}^2\tilde{\epsilon}_{ss}^m+s_{24}^2\epsilon_{\mu\mu}^m) 
\end{array} \right)}\,,
\end{align}
where $R\equiv\delta m_{41}^2/\delta m_{31}^2$, $\hat A = 2\sqrt2 G_F N_e E_\nu/\delta m^2_{31}$, $\tilde{\epsilon}_{ss}^m\equiv \kappa+\epsilon_{ss}^m$, and we have dropped $\delta m^2_{21}$-dependent terms since they are very small. For $R\gg\hat{A}c_{24}s_{24}(\epsilon_{\mu\mu}^m-\tilde{\epsilon}_{ss}^m)$, the measured $\nu_\mu\rightarrow\nu_\mu$ oscillation probability after averaging over the fast oscillation is
\begin{align}
\langle \tilde{P}_{\mu\mu}\rangle \approx (1+2\epsilon_{\mu\mu}^{D}-2s_{24}^2)(1-\sin^22\tilde{\theta}_{23}\sin^2\tilde{\Delta}_{31})\,,
\label{eq:prob1}
\end{align}
where 
$
\sin^22\tilde{\theta}_{23}=\frac{\sin^22\theta_{23}}{C}
$, $
\tilde\Delta_{31}=\frac{\delta m_{31}^2L}{4E_\nu}\sqrt{C}.
$
and $C=\sin^22\theta_{23}+[\cos2\theta_{23}-\hat{A}(c_{24}^2\epsilon_{\mu\mu}^m-\epsilon_{\tau\tau}^m+s_{24}^2\tilde{\epsilon}_{ss}^m)]^2$.
If 
\be
\epsilon_{\mu\mu}^{D} \simeq s_{24}^2\,,
\label{eq:epsdmm}
\ee
and 
\begin{align}
\epsilon_{\mu\mu}^m-\epsilon_{\tau\tau}^m
\simeq s_{24}^2(\epsilon_{\mu\mu}^m-{\epsilon}_{ss}^m- \kappa)\,,
\label{eq:deg}
\end{align}
the measured $\nu_\mu\rightarrow\nu_\mu$ oscillation probability reduces to the standard three-neutrino result.

\begin{figure}
	\includegraphics[width=0.53\textwidth]{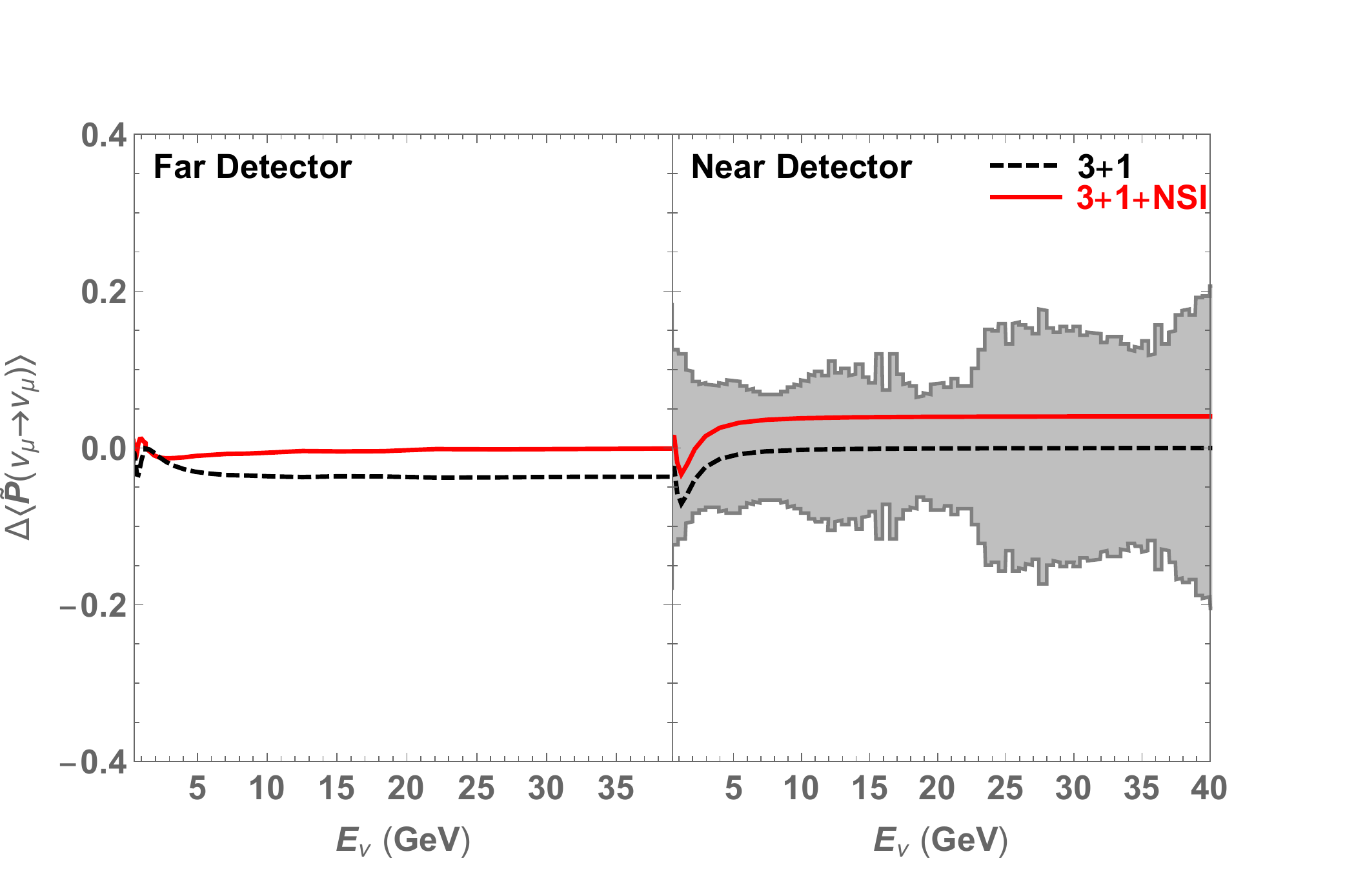}
	\caption{The difference of the measured oscillation probabilities between the 3+1 and standard three-neutrino oscillation cases at the MINOS+ far detector (left) and near detector (right). The solid (dashed) curve corresponds to the case with (without) NSI. Here $\delta m_{41}^2=1.4\text{ eV}^2$, $\sin^2 \theta_{14}=\sin^2 \theta_{14}=\epsilon_{\mu\mu}^D=0.02$, $\epsilon_{\mu\mu}^{m}=-0.7$, $\epsilon_{\tau\tau}^{m}=-0.5$, and $\epsilon_{ss}^{m}=6$, and the other mixing angles and mass-squared differences are the best-fit values in Ref.~\cite{Esteban:2016qun}. The fast oscillations have been averaged out. The shaded band represents the $1\sigma$ systematic uncertainties at the near detector. }
	\label{fig:prob}
\end{figure}

In the rest of the paper we choose the following parameter set to demonstrate consistency with various data:
\begin{eqnarray}
\delta m_{41}^2&=&1.4\text{ eV}^2\,, \ \ \  \sin^2 \theta_{14}=\sin^2 \theta_{24}=\epsilon_{\mu\mu}^D=0.02\,,\nonumber \\
 \epsilon_{\mu\mu}^{m}&=&-0.7\,,\ \ \  \epsilon_{\tau\tau}^{m}=-0.5\,,\ \ \  \epsilon_{ss}^{m}=6\,.
 \label{point}
\end{eqnarray}
In the left panel of Fig.~\ref{fig:prob}, we plot the difference of the measured oscillation probabilities between the sterile and $3\nu$ cases at the MINOS+ far detector (FD) after averaging out the fast oscillations. We see that in the presence of CC NSI, the measured oscillation probability at the MINOS+ FD is almost the same as the standard case, so using the MINOS+ FD data alone cannot distinguish the 3+1 case from the standard three-neutrino oscillation case.  

\begin{figure}
	\includegraphics[width=0.45\textwidth]{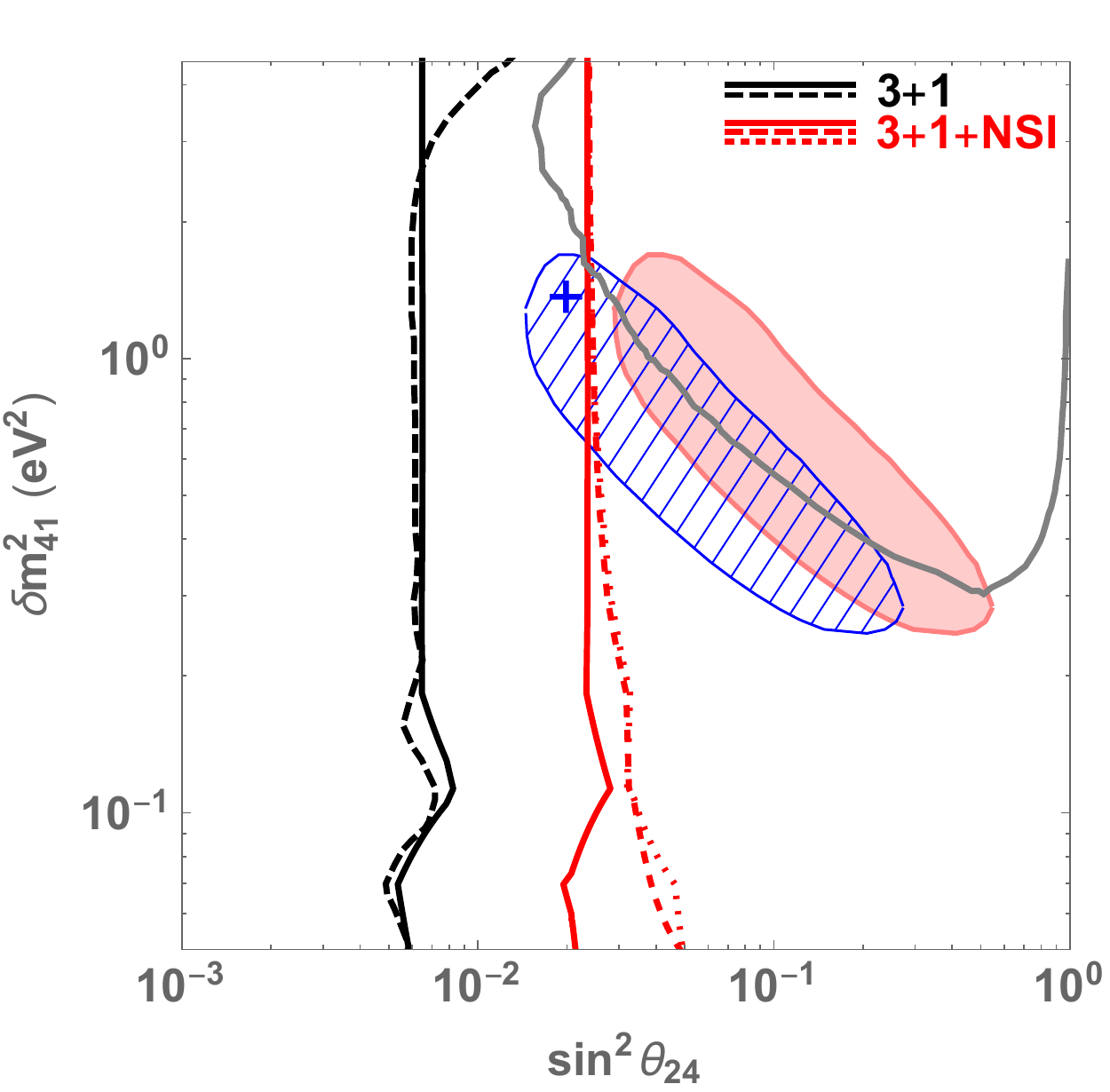}
	\caption{The 90\%~C.L. exclusion limits for the 3+1 scenario from MINOS and MINOS+ data. The dashed black curve is extracted from Ref.~\cite{Adamson:2017uda}, and the black solid curve corresponds to the 3+1 case from our analysis of the FD data only. The red curves correspond to the 3+1+NSI case with $\epsilon_{\mu\mu}^{D}=0.02$. The solid red curve corresponds to no NC NSI, while the dashed red curve corresponds to $\epsilon_{\mu\mu}^{m}=-4.3$ and $\epsilon_{\tau\tau}^{m}=-4$, and the dotted red curve corresponds to $\epsilon_{\mu\mu}^{m}=-0.7$, $\epsilon_{\tau\tau}^{m}=-0.5$ and $\epsilon_{ss}^{m}=6$. The shaded (hatched) region corresponds to the $3\sigma$ allowed region for the combined LSND and MiniBooNE appearance analysis~\cite{Dentler:2018sju} with $\sin^2\theta_{14}=0.01$ ($0.02$). The gray curve corresponds to the CDHS 90\% C.L. exclusion limit, as shown in Ref.~\cite{Adamson:2017uda}. The blue plus sign marks the point in Eq.~(\ref{point}).
	}
	\label{fig:chis}
\end{figure}

{\bf MINOS/MINOS+ analysis.}
To analyze the MINOS and MINOS+ data, we follow the procedure described in Ref.~\cite{Adamson:2017uda}, with the $\chi^2$ defined as 
\be
\chisq =\sum_{i=1}^{71} \sum_{j=1}^{71} (x_i-\mu_i)[V^{-1}]_{ij}(x_j-\mu_j) \,,
\ee
where $x_i$ ($\mu_i$) are the number of observed (predicted) events at the FD, and the covariance matrix $V$ is taken from the ancillary files of Ref.~\cite{Adamson:2017uda}. We modified the oscillation probabilities in the code provided in the ancillary files of Ref.~\cite{Adamson:2017uda} by using the GLoBES software~\cite{GLOBES}, which includes the new physics tools developed in Ref.~\cite{Kopp:2006wp}. In our analysis, we only use the FD data for two reasons: (i) for the mass-squared difference relevant to LSND/MiniBooNE, the sensitivity to constrain sterile neutrinos at MINOS/MINOS+ mainly comes from the FD since the oscillation effects at the MINOS/MINOS+ near detector (ND) are negligible, and (ii) the systematic uncertainties at the ND are very large (see Fig.~\ref{fig:prob}) and a precise determination of the spectrum at the ND has been called into question~\cite{Louis:2018yeg}.

Since the MINOS/MINOS+ data are not sensitive to $\theta_{14}$, we fix $\sin^2\theta_{14}=0.02$.  Hence, the $\chisq$ function for the 3+1 scenario with CC NSI depends only on $\sin^2 \theta_{23}$, $\delta m^2_{23}$, $\sin^2 \theta_{24}$, $\delta m^2_{41}$, and the NSI parameters. For a fixed set of NSI parameters, we marginalize over  $\sin^2 \theta_{23}$ and $\delta m^2_{23}$ for each point in the $(\sin^2 \theta_{24},\delta m^2_{41})$ plane, and calculate $\Delta \chisq (\sin^2 \theta_{24},\delta m^2_{41}) = \chisq_\text{min}(\sin^2 \theta_{24},\delta m^2_{41})-\chisq_\text{min,$3 \nu$}$ to obtain the exclusion limits on the 3+1 model. The resulting $\chisq_\text{min,$3 \nu$}=74.8$ represents a good fit to the 71 data points used in our analysis.

The 90\% C.L. exclusion limits in the ($\sin^2\theta_{24}$, $\delta m_{41}^2$) plane for $\epsilon_{\mu\mu}^{D}=0.02$ are shown in Fig.~\ref{fig:chis}. The dashed black curve is extracted from Ref.~\cite{Adamson:2017uda} and was obtained from an analysis of both the ND and FD data, and the solid black curve corresponds to the 3+1 case from our analysis of the FD data only; clearly, the limits are in good agreement for $\delta m_{41}^2<3 \text{ eV}^2$. The red curve corresponds to the NSI cases with $\epsilon_{\mu\mu}^{D}=0.02$. (Note that the current bounds on vector-like $\epsilon_{\mu\mu}^{ud}$ are rather weak~\cite{Biggio:2009nt}.)
The limits can be understood from Eq.~(\ref{eq:epsdmm}). In general, the bounds become weaker as $\epsilon_{\mu\mu}^{D}$ is  increased. For $\sin^2\theta_{14}=0.01$, the LSND/MiniBooNE allowed region is consistent with the MINOS/MINOS+ data for $\epsilon_{\mu\mu}^D>0.03$. Since larger values of $\theta_{14}$ require correspondingly smaller values of $\theta_{24}$ to explain the LSND/MiniBooNE data, for $\sin^2\theta_{14}=0.02$, large parts of the regions allowed by the appearance data are not constrained by the MINOS/MINOS+ data. From the dashed and dotted red curves we see that NC NSI have a tiny effect on the bounds for eV~scale sterile neutrinos.
Note that CC NSI also increase the number of events at the MINOS/MINOS+ ND. However, these changes are within the systematic uncertainties at the ND~\cite{flux};  see the shaded band in the right panel of Fig.~\ref{fig:prob}. 

{\bf IceCube/DeepCore analysis.}
We now study the atmospheric neutrino constraints in the presence of large NC NSI by combining the IceCube data at high energy and the DeepCore data at low energy. For the IceCube analysis, we follow the procedure of Ref.~\cite{Liao:2016reh}, which analyzed 13 bins in the reconstructed muon energy range, 501~GeV~$\le E_{\mu}^{rec} \le 10$~TeV, and 10~bins in the zenith angle range, $-1 \le \cos\theta_z \le 0$~\cite{Aartsen:2015rwa}.  For the DeepCore analysis, we use the publicly available data from Ref.~\cite{Aartsen:2014yll}, which has 8 bins in the reconstructed energy range, 6~GeV~$\le E_{\mu}^{rec} \le 56$~GeV, and 8~bins in the zenith angle range, $-1 \le \cos\theta_z \le 0$. The expected number of observed events at DeepCore is given by
\bea
&&N_{ij}^{exp} =
\int d\cos\theta_z  \int dE_\nu  \Phi_{\nu_\mu}(E_\nu, \cos\theta_z)P_{\nu_\mu\nu_\mu}(E_\nu, \cos\theta_z)
\nonumber\\
&&  \times A_{eff} (E_{\mu, i}^{rec}, \cos\theta_{z, j}, E_\nu, \cos\theta_z)+(\nu\rightarrow\bar{\nu}) \,,
\label{eq:Nexpij}
\eea
where $\cos \theta_z$ is the cosine of the zenith angle, $\Phi_{\nu_\mu}(E_\nu, \cos\theta_z)$ is the atmospheric $\nu_\mu$ flux at the surface of the earth~\cite{Honda:2015fha}, $P_{\nu_\mu\nu_\mu}(E_\nu, \cos\theta_z)$ is the $\nu_\mu\rightarrow\nu_\mu$ oscillation probability at the detector, and $A_{eff}$ is the neutrino effective area given in Ref.~\cite{Aartsen:2014yll}.

\begin{figure}
	\includegraphics[width=0.45\textwidth]{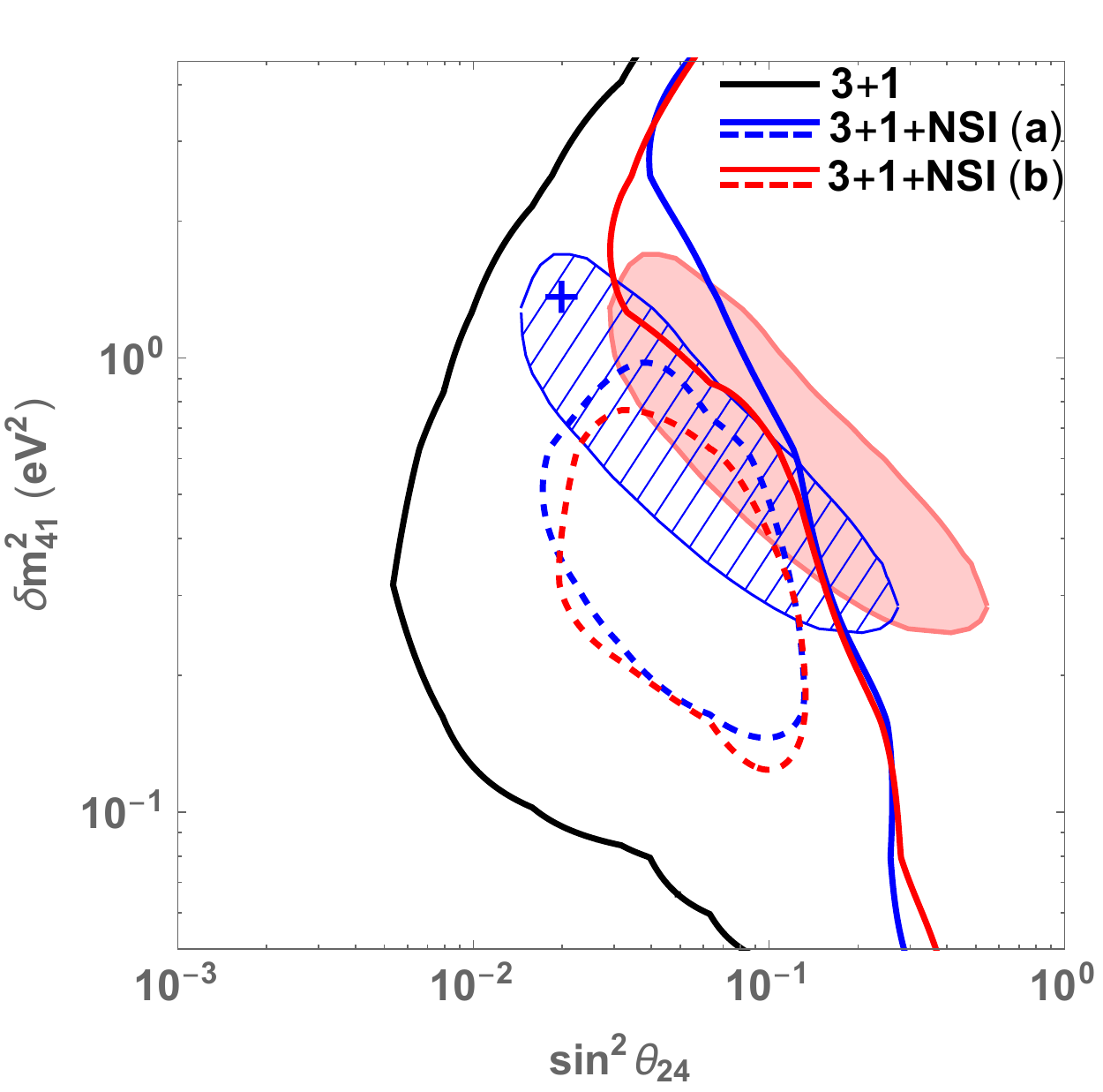}
	\caption{The 90\% C.L. exclusion limits for the 3+1 scenario from IceCube and DeepCore data. The solid black curve corresponds to the 3+1 oscillations without NSI, and the solid red [blue] curve corresponds to the 3+1+NSI (a) [(b)] case. The red (blue) dashed contour corresponds to the 90\% C.L. allowed region in case (a) [(b)]. The shaded (hatched) region corresponds to the $3\sigma$ allowed region for the combined LSND and MiniBooNE appearance analysis~\cite{Dentler:2018sju} with $\sin^2\theta_{14}=0.01$ ($0.02$). $\epsilon_{\mu\mu}^{D}=0.02$ for the NSI scenarios. The blue plus sign marks the point in Eq.~(\ref{point}).}
	\label{fig:icdc}
\end{figure}

To calculate the statistical significance of an oscillation scenario, we define 
\begin{align}
\chisq_\text{DC} &=2\sum_{i,j=1}^{8}\left[ N_{ij}^{th}(\alpha,\beta) - N_{ij}^{obs} + N_{ij}^{obs}\ln\frac{N_{ij}^{obs}}{ N_{ij}^{th}(\alpha,\beta)}\right]
\nonumber\\
&+ \frac{(1-\alpha)^2}{\sigma_\alpha^2}
 \,,
\end{align}
where $N_{ij}^{obs}$ is the observed event counts per bin,  and $N_{ij}^{th}(\alpha,\beta)=\alpha N_{ij}^{exp}+\beta N_{ij}^{bkg}$ with $N_{ij}^{bkg}$ being the atmospheric muon background per bin. 
We take the uncertainty in the atmospheric neutrino flux normalization to be $\sigma_\alpha=20\%$ at the energies relevant to DeepCore, and we allow the normalization of the atmospheric muon background to float freely~\cite{Aartsen:2014yll}. We find 
       $\chisq_\text{DC,min,$3\nu$}=60.7$ with $\alpha=0.869$ and $\beta=0.184$, and confirmed that the confidence regions for the standard $3\nu$ oscillation from our analysis agree with those in Ref.~\cite{Aartsen:2014yll}.

\begin{table}[t]
	\begin{center}
		\begin{tabular}{|c|c|c|c|c|c|c|c|c|}
			\hline
			Case & $\epsilon_{\mu\mu}^m$ &  $\epsilon_{\tau\tau}^m$ &  $\epsilon_{ss}^m$ &  $\sin^2\theta_{24}$ & $\delta m^2_{41}$  &   $\chisq_\text{IC} $ &   $\chisq_\text{DC} $ &   $\chisq_\text{IC+DC,min} $ \\
			\hline
			(a) & -4.3 & -4 & 0 & 0.063 & 0.32 & 103.4 & 54.9 & 158.3  \\
			\hline
			(b) & -0.7 & -0.5 & 6 & 0.032 & 0.63 & 102.7 & 56.5 & 159.2 \\
			\hline
		\end{tabular}
	\end{center}
	\label{tab:min}
	\caption{The minimum $\chi^2_\text{IC+DC}$ for the best-fit scenarios. In case (a), the scanned parameter ranges are $|\epsilon_{\tau\tau}^m|<6$ and $|\epsilon_{\mu\mu}^m-\epsilon_{\tau\tau}^m|<0.5$; in case (b), the scanned parameter ranges are $|\epsilon_{ss}^m|<6$, $|\epsilon_{\tau\tau}^m|<0.5$ and $|\epsilon_{\mu\mu}^m-\epsilon_{\tau\tau}^m|<0.5$. There are 130~IceCube and 64~DeepCore data points in the analysis.}
\end{table}

To obtain the exclusion regions for the combined IceCube and DeepCore data in the 3+1 scenarios, we calculate $\Delta \chisq_\text{IC+DC} (\sin^2 \theta_{24},\delta m^2_{41})= \chisq_\text{IC+DC,min}(\sin^2 \theta_{24},\delta m^2_{41})-\chisq_\text{IC+DC,min,$3\nu$}$, where $\chisq_{\text{IC+DC}}=\chisq_\text{IC}+\chisq_\text{DC}$ for each set of parameters;
$\chisq_\text{IC+DC,min,$3\nu$}=172.7$. The exclusion region for the 3+1 scenario without NSI is shown as the black line in Fig.~\ref{fig:icdc}. We see that the LSND/MiniBooNE allowed region is excluded by the IceCube/DeepCore data.

We consider two NSI cases: (a) only $\epsilon_{\mu\mu}^m$ and $\epsilon_{\tau\tau}^m$ are nonzero, and (b) $\epsilon_{\mu\mu}^m$, $\epsilon_{\tau\tau}^m$, and $\epsilon_{ss}^m$ are all nonzero. For case (a) we scan the parameter space, $|\epsilon_{\tau\tau}^m|<6$ and $|\epsilon_{\mu\mu}^m-\epsilon_{\tau\tau}^m|<0.5$. 
For case (b), we scan the parameter space, $|\epsilon_{ss}^m|<6$, $|\epsilon_{\tau\tau}^m|<0.5$ and $|\epsilon_{\mu\mu}^m-\epsilon_{\tau\tau}^m|<0.5$. We note that large $\epsilon_{\mu\mu}^m$ and $\epsilon_{ee}^m=0$ yield large $\epsilon_{\mu\mu}^m-\epsilon_{ee}^m$, which may be constrained by solar data~\cite{Esteban:2018ppq}. Therefore, in order to accommodate a small value for $\epsilon_{\mu\mu}^m-\epsilon_{ee}^m$, we allow large $\epsilon_{ss}^m$ in case (b). (The global analysis of Ref.~\cite{Esteban:2018ppq} uses solar data to place constraints on $\epsilon_{\mu\mu}^m-\epsilon_{ee}^m$, which however do not apply to our scenario because it includes a sterile neutrino.)
We also fixed 
       $\epsilon_{\mu\mu}^{D}=0.02$ for both cases.
       Our results are not sensitive to the value of $\epsilon_{\mu\mu}^{D}$, since $\epsilon_{\mu\mu}^{D}$ only affects the overall normalization of the expected events and the uncertainty of the atmospheric neutrino flux normalization is large. The minimum values of $\chi^2_\text{IC+DC}$ for both cases are given in Table~\ref{tab:min}. We find an allowed region that is consistent with the LSND and MiniBooNE data for each case. The best-fit parameters in both cases are consistent with Eq.~(\ref{eq:deg}).
The exclusion regions for the 3+1 scenario in the presence of NSI are shown as the blue [red] lines in Fig.~\ref{fig:icdc} for case (a) [(b)].  We see that the LSND and MiniBooNE allowed region is consistent with IceCube/DeepCore data in the presence of large NC NSI in the active neutrino sector, or large NC NSI in the sterile neutrino sector and small NC NSI in the active neutrino sector.

\begin{figure}[t]
	\includegraphics[width=0.52\textwidth]{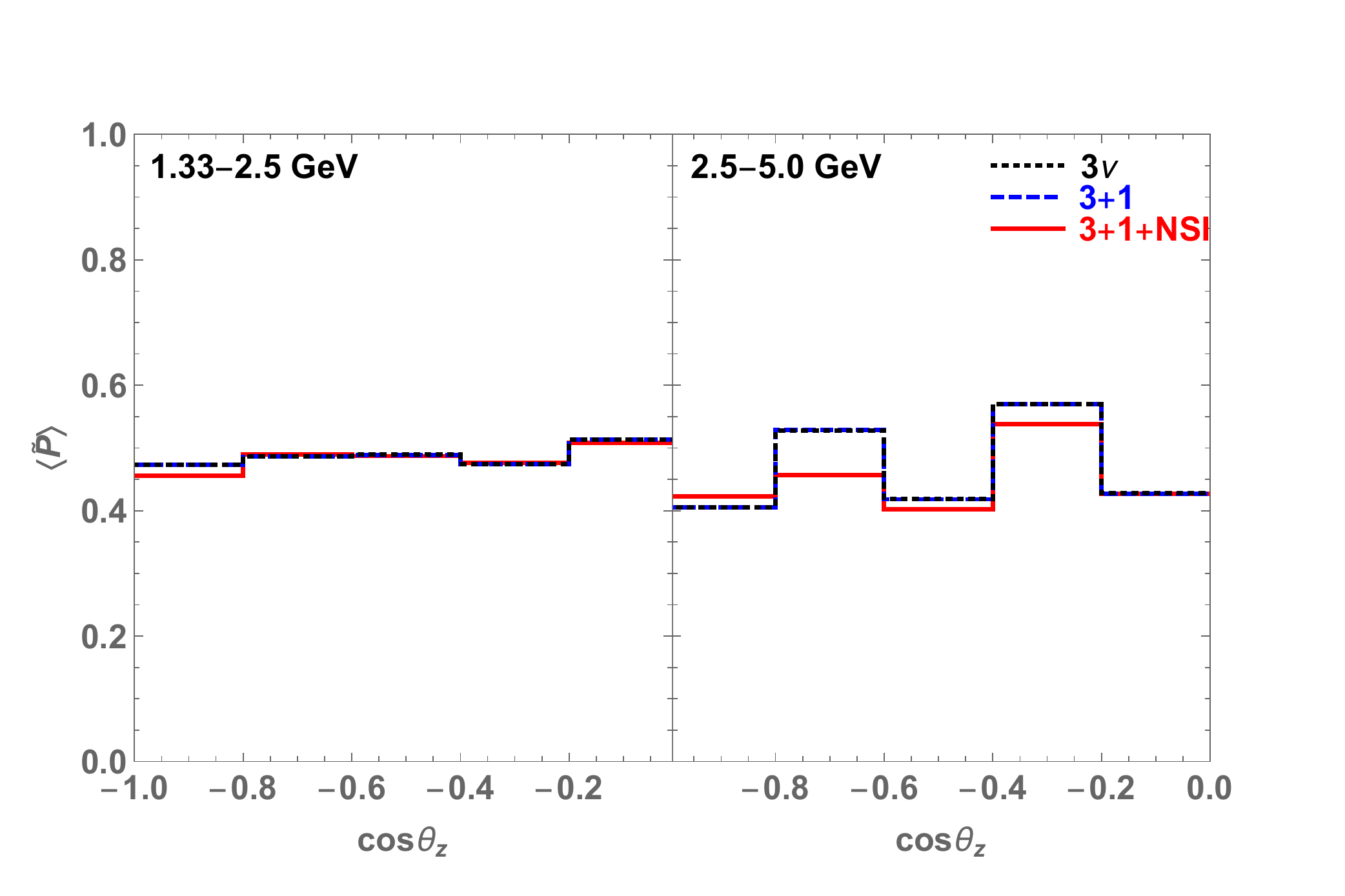}
	\caption{Zenith angle distributions of the averaged atmospheric muon neutrino and antineutrino survival probabilities for two Super-Kamiokande energy bins. The solid (dashed) [dotted] curve corresponds to the 3+1+NSI (3+1) [3$\nu$] case. For the 3+1 case without NSI, the spectra are normalized by a factor of 1.04 to compare with the $3\nu$ case. The oscillation parameters are the same as in Fig.~\ref{fig:prob}. }
	\label{fig:sk}
\end{figure}

{\bf Other data.} The Super-Kamiokande (SK) experiment has collected atmospheric neutrino events with energies lower than at DeepCore. We checked that the differences of the survival probabilities between the NSI and $3\nu$ cases are within the statistical uncertainties for two SK multi-GeV energy bins~\cite{sk}; see Fig.~\ref{fig:sk}. For the sub-GeV events at SK, systematic uncertainties are very large due to the poor angular correlation between the neutrino and outgoing lepton~\cite{Kajita:2016vhj}.

\begin{figure}[t]
	\includegraphics[width=0.45\textwidth]{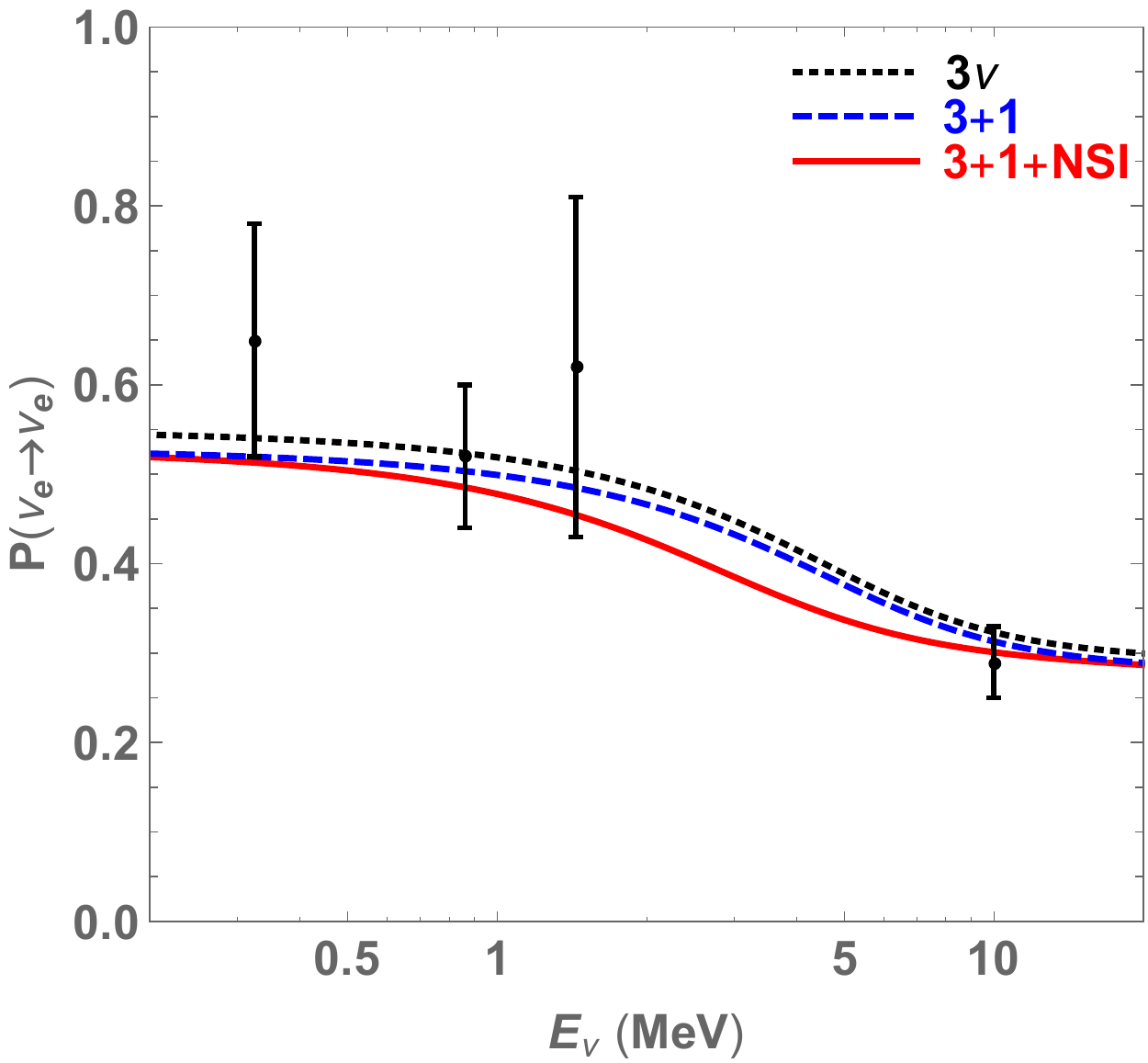}
	\caption{The survival probabilities of solar neutrinos for the $3\nu$, 3+1, and 3+1+NSI cases. The oscillation parameters are the same as in Fig.~\ref{fig:prob}.  }
	\label{fig:solar}
\end{figure}

Solar neutrino propagation is sensitive to modifications of the matter potential. To analyze solar neutrino data, we follow the procedure of Ref.~\cite{Barger:2005si} in conjunction with the Standard Solar Model fluxes~\cite{Vinyoles:2016djt}. The survival probabilities obtained from the Borexino measurements of the $pp$~\cite{Bellini:2014uqa}, $^7$Be~\cite{Bellini:2011rx}, and $pep$ neutrinos~\cite{Collaboration:2011nga}, and the SNO CC measurement of the high energy ($^8$B and $hep$) neutrinos~\cite{Aharmim:2008kc}, are the four data points in Fig.~\ref{fig:solar}.  The survival probabilities for the $3\nu$ and NSI cases are also shown. We find  $\chi^2=1.79$ and 2.13 for the $3\nu$ and NSI case, respectively, demonstrating compatibility of the 3+1+NSI scenario with current solar data.  Note that since NSI shift the upturn in the survival probability to
lower energies, the tension between KamLAND and $^8$B neutrino data is eased.

We mention in passing that data from the appearance channels at current long-baseline experiments cannot distinguish between the $3\nu$ and NSI cases. 

{\bf Summary.} 
       MINOS+ and IceCube data present a challenge to an explanation of the LSND/MiniBooNE anomaly with the simple 3+1 model.  If the measurements of these experiments are accepted prima facie, the 3+1 model must be extended by introducing baroque new physics to make the data compatible with each other. We find that effects of the sterile neutrino at MINOS+ can be canceled by CC NSI at the detector via 
       $\epsilon_{\mu\mu}^{D}=2\epsilon_{\mu\mu}^{udL}$, thereby significantly weakening the MINOS+ constraint on the sterile neutrino parameter space. Also, the
 LSND/MiniBooNE allowed regions can be made consistent with IceCube and DeepCore data by including large matter NSI parameters, 
 $\epsilon_{\mu\mu}^m$ and  $\epsilon_{\tau\tau}^m$, or large  
       $\epsilon_{ss}^m$ and small $\epsilon_{\mu\mu}^m$ and $\epsilon_{\tau\tau}^m$. The CC and NC NSI parameter values required do not impact the data taken by the MiniBooNE and LSND experiments. A global fit of the 3+1+NSI
scenario is needed to conclusively confirm our findings.
       
The CC NSI parameter $\epsilon_{\mu\mu}^{udL}$ can be directly constrained at the DUNE~\cite{Bakhti:2016gic} and MOMENT~\cite{Tang:2017qen} experiments. Also, large diagonal NC NSI will lead to a modification of the matter potential, which will be tested at future long-baseline experiments~\cite{Liao:2016orc}. 
A study of early universe cosmology in the 3+1 scenario with CC and NC NSI is underway by a subset of us.

{\it Acknowledgments.} We thank A.~Aurisano, and J.~Todd for helpful correspondence. K.W. thanks the University of Hawaii for its hospitality while this work was in progress. This research was supported in part by the
U.S. DOE under Grant No. DE-SC0010504.

\vspace{0.1 in}

\vskip1cm

\newpage

\end{document}